\documentclass[conference, draftcls, onecolumn]{IEEEtran}
\IEEEoverridecommandlockouts
\usepackage{amsmath,amssymb,amsfonts}
\usepackage{algorithmic}
\usepackage{graphicx}
\usepackage{textcomp}
\usepackage{xcolor}

\usepackage{cite} 

\begin{document}

\title{COSMICA: a novel parallel GPU code for Cosmic Rays propagation in heliosphere}

\author{\IEEEauthorblockN{1\textsuperscript{st} Giovanni Cavallotto}
\IEEEauthorblockA{\textit{National Institute for Nuclear Physics (INFN)}, \\
\textit{Division of Milano-Bicocca;} \\
\textit{ICSC - Centro Nazionale di Ricerca in HPC,} \\
\textit{Big Data and Quantum Computing} \\
giovanni.cavallotto@mib.infn.it}
\and
\IEEEauthorblockN{2\textsuperscript{nd} Stefano Della Torre}
\IEEEauthorblockA{\textit{National Institute for Nuclear Physics (INFN)}, \\
\textit{Division of Milano-Bicocca;} \\
\textit{ICSC - Centro Nazionale di Ricerca in HPC,} \\
\textit{Big Data and Quantum Computing} \\
stefano.dellatorre@mib.infn.it}
\and
\IEEEauthorblockN{3\textsuperscript{rd} Giuseppe La Vacca}
\IEEEauthorblockA{\textit{University of Milano-Bicocca,} \\
\textit{Department of Physics ``Giuseppe Occhialini''} \\
giuseppe.lavacca@mib.infn.it}
\and
\IEEEauthorblockN{4\textsuperscript{th} Massimo Gervasi}
\IEEEauthorblockA{\textit{University of Milano-Bicocca,} \\
\textit{Department of Physics ``Giuseppe Occhialini''} \\
massimo.gervasi@mib.infn.it}
}

\maketitle

\begin{abstract}
The complex structure of interplanetary magnetic fields and their variability, due to solar activity, make it necessary to compute the Cosmic Ray (CR) modulation with numerical simulations.
COde for a Speedy Monte Carlo (MC) Involving Cuda Architecture (COSMICA) is a MC code, solving backward-in-time the system of Stochastic Differential Equations (SDE) equivalent to the Parker Transport Equation (PTE).
The Graphics Processing Unit (GPU) parallelization of COSMICA code is a game changer in this field because it reduces the computational time of a standard simulation from the order of hundred of minutes to few of them. Furthermore, the code is capable of distributing the computations on clusters of machines with multiple GPUs, opening the way for scaling. In COSMICA we implemented the synchronous broadcasting of memory access for evolving variable samples, the rounding of virtual particle set numbers, to fulfil the GPU blocks, and the exploitation of shared memory to free registers. Furthermore, we compactify the mathematical computations and pass to the lighter momentum formulation of SDE. The first porting of the code on GPU architecture brings it to a speed-up of 40X. The successful optimizations bring 1.5X speed-up.
\end{abstract}


\newpage

\section{Introduction}

Galactic Cosmic Rays (GCR) are charged particles mainly originating from astrophysical sources outside the solar system. These particles enter isotropically the heliosphere, which is the region dominated by the Solar Wind (SW) and its magnetic fields. As they traverse it, GCRs undergo a diffusion dominated transport process, which is related to solar activity and cause a time variable flux, that is known as "solar modulation of CRs". Understanding GCR modulation is crucial for fundamental physics, offering insights into the interplay between cosmic rays and the SW plasma inside our solar system's environment. Moreover, it gives insights about their galactic sources and their production processes. On the other hand, it has direct applications for space exploration, where CRs pose a significant risk to spacecraft electronics and astronaut health. Developing accurate and efficient CR propagation simulation tools is therefore essential for advancing our knowledge in this field and mitigating potential risks for future space missions.
The transport of charged particles in a turbulent magnetic field is described by the Parker Transport Equation (PTE), which is a Fokker-Plank-like equation, incorporating the diffusion, convection, drift, and adiabatic energy loss processes.

Classical techniques for solving multi-dimensional partial differential equations often rely on numerical integration schemes, such as finite difference methods (see, e.g., \cite{Jokipii_1979, Kota_1983, Potgieter_1985, Burger_1995}) or implicit difference methods (see, e.g., \cite{Fisk_1971, Kóta_1991}). However, these approaches are prone to significant challenges, particularly numerical instabilities when applied to higher-dimensional problems \cite{Pei_2010, KOPP2012530}.
The state-of-the-art approach to simulating GCR propagation, both in terms of accuracy and computational efficiency, is to solve, instead, a system of Stochastic Differential Equation (SDE) equivalent to PTE.

Actually, there are few numerical codes used to study these phenomena and among the most advanced, one can find HelMod \cite{Boschini_2018_a, BOSCHINI20192459}, Geliosphere \cite{SOLANIK2023108847}, SOLARPROP \cite{Kappl_2016, Fiandrini_2021}, SDEMMA \cite{Qin_2017, Shen_2019, Song_2021} codes and other codes developed for specific studies \cite{Strauss_2011_a, Strauss_2011_b, MOLOTO2019, DUNZLAFF2015156}. There are a few parallelized adaptations of them (see e.g \cite{SOLANIK2023108847, DUNZLAFF2015156, MOLOTO2019}), which are successful in speeding up their simulations. Nevertheless, these codes are limited by the fact that they are not natively designed to run on High Performance Computing (HPC) infrastructure. We developed a new generation of Heliospheric propagation code, named COSMICA, which is natively built upon GPU architecture, leveraging the cuda low-level programming and that embeds the most recent fine structure observations of the heliosphere. Furthermore, COSMICA will be open source and customizable with various physical models.

\section{Solar modulation of GCRs}

Modelling GCR modulation requires of various components of the whole physical model, which are illustrated in the followings.

\subsection{Parker Equation}

GCR are charged particles travelling in a low density quasi-ideal plasma medium, filled with turbulent magnetic fields. Therefore, their complete dynamics is collected inside the Parker Transport Equation (PTE), named after Eugene Parker who first proposed it in the 1965 (see, e.g., \cite{parker1965, BOSCHINI20192459} and references therein):
\begin{align}
\label{eq:ParkerEquation}
\frac{\partial U}{\partial t}= &\frac{\partial}{\partial x_i} \left( K^S_{ij}\frac{\partial \mathrm{U} }{\partial x_j}\right) +\frac{1}{3}\frac{\partial V_{ \mathrm{sw},i} }{\partial x_i} \frac{\partial }{\partial T}\left(\alpha_{\mathrm{rel} }T\mathrm{U} \right) \\
&- \frac{\partial}{\partial x_i} [ (V_{ \mathrm{sw},i}+v_{d,i})\mathrm{U}], \notag
\end{align}

where $U$ is the number density of GCR particles per unit of kinetic energy $T$ (GeV/nucleon), $t$ is time, $V_{ \mathrm{sw},i}$ is the SW velocity along the axis $x_i$, $K^S_{ij}$ is the symmetric part of the diffusion tensor, $v_{d,i}$ is the particle magnetic drift velocity (related to the antisymmetric part of the diffusion tensor), and $\alpha_{\mathrm{rel} }=\frac{T+2m_r c^2}{T+m_r c^2}$, with $m_r$ the particle rest mass per nucleon in units of GeV/nucleon.
The various terms inside the PTE corresponds to physical processes affecting GCR along their path and described in many papers (see, e.g., \cite{Boschini_2018_a, BOSCHINI20192459, Qin_2017}). 

These phenomena affect GCR locally and are time-dependent, governed by the solar activity and dynamics, which evolve and are non-linear. Additionally, the PTE is in function of particle rigidity and charge sign, producing different modulation for different CR species, for different solar polarity.

\subsection{The physical model}


In this work, we refer to the HelMod model as the underlying physical model of the code. The choice of this model is motivated by the fact that it has a long development story \cite{Boschini_2017, Boschini_2018_a, Boschini_2018_b, BOSCHINI20192459, Boschini_2020, BOSCHINI20222649, BOSCHINI20244302} and it has been validated by independent studies \cite{Bartocci_2020, Rankin_2022, Liu_2024}. Its parameter values are tuned by means of a fitting procedure on Alpha Magnetic Spectrometer (AMS-02) proton and Helium daily flux~\cite{2021PhRvL.127A1102A, 2022PhRvL.128w1102A}.
There are more complete models considering magnetic hydrodynamical simulations of the plasma \cite{Izmodenov_2015, Izmodenov_2020, Michael_2021, Michael_2022}, which however are computational prohibitive.
The simulation of the whole structure of the heliosphere (see, e.g., \cite{Kleimann_2022}) is a critical aspect of the GPU implementation. In this work, we divide the heliosphere into two regions: inner heliosphere and heliosheath, both squashed in the nose direction (i.e. the direction in which the sun is travelling with respect to the interstellar medium frame). Furthermore, the inner heliosphere is divided into 15 expanding shells corresponding to the regions averaged on Carrington rotation (i.e., a synodic rotation period of 27.26 days). Each region contains different propagation parameters that should be stored on GPU memory and accessible to all threads.

The final ingredient to simulate the CR heliospheric modulation is the Local Interstellar Spectrum (LIS), which is the spectrum of GCRs outside the heliosphere. Its knowledge is mandatory to estimate the fluxes measured inside the heliosphere, after the modulation. In this paper, we use the Galprop LIS \cite{Boschini_2017} up to 28 ion species, tuned along with HelMod Code on the most updated set of CR measurements \cite{Boschini_2017}.

\section{Computational method}

The PTE is a Fokker-Planck type equation that can be solved using both a forward-in-time and backward-in-time approach (in the latter case it is usually named as Kolmogorov equation, see Equation 1.7.15 in \cite{kloeden1999}). Both Fokker-Planck and Kolmogorov differential equations are equivalent to a system of SDE, as shown in Sections 4.3.2–4.3.5 of \cite{gardiner1985} and Appendix A.13.1 of \cite{HandBookMC}, following the Ito's formula. Forward-in-time approach traces particle-like objects from the heliosphere boundary toward the target, while is the opposite for the backward-in-time one. The latter methodology, is lighter in computational terms because only the particles actually reaching the target are simulated avoiding filling the heliosphere with errant particles, as discussed in details in \cite{DellaTorre2016_OneD}.

To obtain the backward-in-time SDEs equivalent to \eqref{eq:ParkerEquation}, the latter should be rearranged to match the following formulation (see, e.g., Equation 13 of \cite{Zhang1999}, Equation A2 of \cite{Strauss_2011_a}, Equation 14 of \cite{Kopp2012}):
\begin{align}
\label{eq:fokkerPlanck}
 \frac{\partial Q}{\partial s} &= \sum_i A_{i}(s,y) \frac{\partial Q}{\partial y_i}+\frac{1}{2}\sum_{i,j} C_{ij}(s,y) \frac{\partial^2 Q}{\partial y_i\partial y_j} \\
 & \quad -L_B Q + S \notag
\end{align}
where $Q$ represent the evolving quantity, $A_{B,i}$ is the advective vector, $C_{B,ij}$ is the diffusion tensor, $L_B$ describes energy loss and $S$ stances for source of particles. In this formulation, $\partial s>0$ represents the backward time evolution of the propagation. The system of SDEs corresponding to \eqref{eq:fokkerPlanck} can be generally expressed as:
\begin{equation}\label{eq:SDE}
 dy_i(s) = A_{i}ds+B_{i,j}dW_j(s),
\end{equation}
where tensors $B$ and $C$ follow the relationship $C= B B^T $, and $d\vec W$ represents the increments of a \emph{standard Wiener process}, which can be described as an independent random variable of the form  $\sqrt{d s} N(0,1)$, with $N(0,1)$ denoting a normally distributed random variable with zero mean and unit variance (see, e.g., Appendix A of \cite{Zhang1999} and Section 2 of \cite{Higham2001}). To numerically integrate the SDEs, the Euler-Maruyana scheme (see, e.g., \cite{HandBookMC}, Section 5.6.1) is the simplest and most commonly used, combined with the Ito rule (see discussion in \cite{DellaTorre2016_OneD} and reference therein). For the complete derivation of each component of $A, B$ and $C$ see appendix A.1 and A.2 of \cite{Boschini_2018_a}.
The backward-in-time approach, despite a less trivial physical representation, has the advantage to simulate only the \textit{quasi-particles}, actually reaching the detection point along their path inside the heliosphere. In the forward-in-time approach, instead, one has to select the \textit{quasi-particles} hitting the detection point from the swarm of them, which is a small fraction of them (see e.g \cite{DellaTorre2016_OneD}).
Propagating backward-in-time allows one to simulate only the \textit{quasi-particles} actually reaching the restricted subset of phase space points of interest, like the Earth orbit or the spacecraft trajectory (as shown by \cite{Strauss_2017}).

At the end of the backward propagation, one is left with the \textit{quasi-particle} population at the boundary of the heliosphere. This translates into a certain energy distribution for each particle species. Thus, the final spectra are modulated with a weight corresponding to the convolution with the LIS at the final energy  and an exponential weight including the \textit{loss} term in \eqref{eq:fokkerPlanck} (see Equation 22 in \cite{DellaTorre2016_OneD}):
\begin{align}
\label{eq:LIS_conv}
    J_{mod}(T_{in}) &\propto \sum^N_{k=1} LIS(T_{ex,k}) \cdot e^{-\sum^k_{j=0} L_{j} \cdot \Delta s}
\end{align}
where $J_{mod}(T_{in})$ is the modulated spectra at energy $T_{in}$, $\mathcal{N}$ is the number of simulated \textit{quasi-particles} with the energy $T_{in}$ at the initial position, $T_{ex,k}$ is the \textit{quasi-particle} energy at heliosphere boundary, $k$ is the number of integration steps from the origin to the boundary, $L_{B,j}$ is the \textit{loss} term in \eqref{eq:fokkerPlanck} evaluated at step $j$, and $\Delta s$ is the integration time step. In the previous lines, the term initial has to be intended as the starting point of the backward propagation (i.e. the ending point of the real particle trajectories).

\section{Code algorithm and implementation}

SDE approach of solving PTE, which reflects also the stochastic nature of CR propagation in turbulent magnetic fields, makes it necessary to simulate the phenomenon with MC methods, which represent a computationally expensive issue. With CPU implementation of the code, the computational power is hardly affordable by a local academic research department. With a GPU parallel approach, instead, the resources allocation and execution time are reduced such that simulations can be generated also for general research studies. Furthermore, this was done using GPUs with competitive prices in comparison to the multicore corresponding CPU clusters.

\subsection{Algorithm structure}\label{sec:algorithm}

The starting point of the COSMICA parallelization is the fact that, to produce the correlation function, between entering end exiting energies, in MC approach a sample of identical \textit{quasi-particles} are let to evolve independently, propagating inside the heliosphere. Then the distribution of this stochastic simulation is taken as code output. Due to the near emptiness of the interplanetary space, no particle-particle interactions occur. Here we are also neglecting the effects of the pressure exerted by CR on the heliospheric plasma, which could modify its shape, especially in the heliosheath \cite{Dorman1998}. Therefore, the temporal evolution of each \textit{quasi-particles} is completely independent and could be easily parallelized on an HPC system, e.g., on GPU architecture. We allocate a GPU thread (i.e., the minimal computing unit of the GPU) for each \textit{quasi-particle}, solving the SDEs at each time step, following the single instruction, multiple data (SIMD) paradigm \cite{CudaCProgramming2014}, which is optimal for GPU architecture.
The code is written in the CUDA-C language\footnote{https://developer.nvidia.com/cuda-toolkit}
(for a complete guide see handbooks in Refs. \cite{CudaCProgramming2014, CudaByExapmple2010, cuda_best_practice_2014}), which provides optimized interactions and low-level code architecture for the NVIDIA GPUs.

\begin{figure}[htb!]
    \centering
    \includegraphics[width=0.8\linewidth]{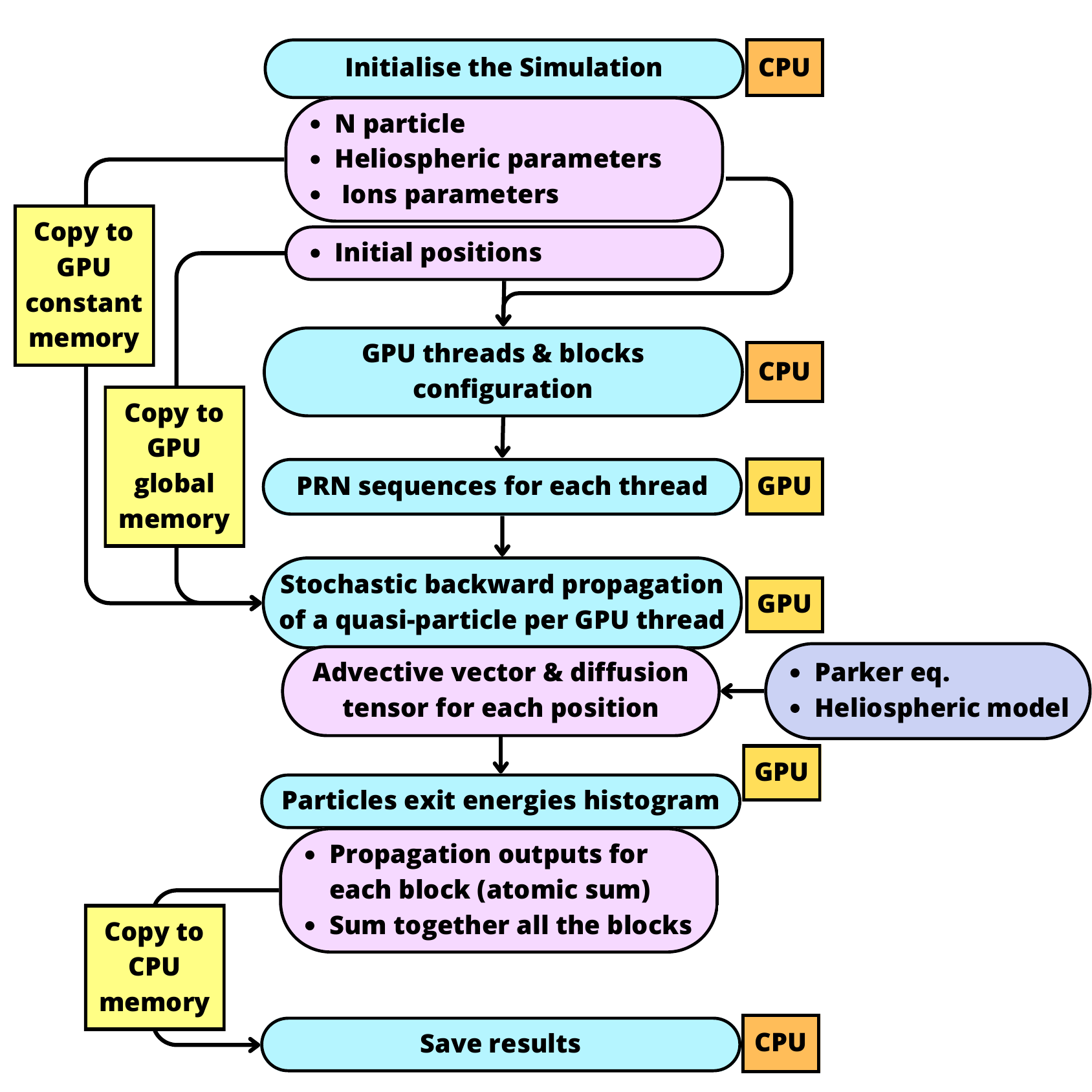}
    \caption{Code-flow chart of COSMICA algorithm. In sky blue are represented the macroscopic steps of the code, in purple the products of the respective step, in yellow the CPU-GPU memory interaction and in orange which processing unit executes the respective code section.}
    \label{fig:GPU_algorithm}
\end{figure}

The code-flow diagram of COSMICA algorithm is shown in Fig.~\ref{fig:GPU_algorithm}. Following it, we encounter first the initialization of the hyperparameters of the simulation: the number of \textit{quasi-particles} to be simulated ($\mathcal{N}$), the heliospheric model parameters, the particle species characteristics, and the initial positions.
Notable it is that COSMICA has incorporated the account for varying detection point position. Allowing to integrate the CRs fluxes over the mission orbit and duration. This is managed, as for a fixed point, by subdividing the \textit{quasi-particles} sample between the set of changing location, following the spacecraft orbit. Common constant variables along the simulation are copied into the constant memory (see, e.g., Chapter 4-5 of \cite{CudaCProgramming2014}), exploiting the read-only protection during the kernel execution. Moreover, the constant memory has short latency, high bandwidth, and, using a broadcasting implemented method, memory reading can be fast as register \cite{cuda_best_practice_2014}. These features make it perfect to store the time-independent and thread-common parameters of the simulation. All the evolving particle coordinates: position, energy and flying time, are stored in global memory during the initialization. Then they are copied to shared memory to reduce the register occupancy and achieve the maximum parallelization. This kind of memory has latency similar to registers. This memory storage choice tends in the direction of relieve register pressure. Indeed, the maximum dimension of shared memory per thread inside a block overwhelms the 32 maximum registers per thread in Ampere architecture GPUs.

The second step in the algorithm is the configuration of the GPU kernel execution and allocation of the hardware resources, where we adjust the number of simulated particle $\mathcal{N}$ to be a multiple of the size of a \textit{warp} (i.e., 32 for NVIDIA GPUs with Compute Capability version 8.0 and 8.6), maximizing the GPU occupancy.
Then we generate the random number sequences needed for the stochastic Wiener process term in \eqref{eq:SDE}, using the \textit{Philox4\_32\_10} generator \cite{Salmon2011}, which is a Pseudo-Random Number Generator (PRNG) provided by the device API of the cuRAND library. One cuda random state is generated once, for each thread, then a random floater is produced for each \textit{quasi-particles} evolution time step. With this implementation, all the threads in the same batch of simulation have the same seed of random sequence, but each individual thread has its own random state. Furthermore, the PRNG is initialized with a different seed for each propagation kernel call (see, e.g., the implementation in \cite{ROMERO2020} and \cite{AskarEtal2021} for the PRNGs choice).

At this point, all is set to execute the stochastic integration of the SDEs independently for each thread. At each integration step cycle, the particles are propagated backward-in-time by one time step, which is set to $50 s$ by default and, eventually, updated to fulfil the physical condition of having the advective term much smaller than the diffusive term \cite{Kruells_1994}. The device computes the particle heliospheric location and the corresponding $A_{B,i}$ , $B_{B,i,j}$ in the \eqref{eq:SDE}, until a heliospheric boundary is reached. Inside the computation of these coefficients is nested the majority of the physical model and assumptions, making them computationally expensive. This is especially true for the register pressure, determined by the complexity of the mathematical expressions involved here. Inside the propagation kernel is computed, also the partial maximum of exiting energies between the block, which is going to be used to build the distribution histogram binning. This is done to leverage the shared locality of partial outputs, avoiding an extra device-host memory transfer. 
Eventually, the exit modulated \textit{quasi-particle} energies are collected in partial histograms, one for each block, with an atomic function to avoid memory conflicts. Then, the latter are merged, and the final results are copied to the host memory. The whole histogram building and filling is executed at device level, with most external memory used to be the global GPU one, avoiding CPU-GPU intermediate memory transfer.

The whole algorithm, illustrated so far, is executed for each initial energy bin initialized, in the frame of backward-in-time integration, corresponding to the energy of the CR when it reaches the Earth, for example. From the code execution point of view, each initial energy bin is assigned to a CPU-GPU couple using omp pragma API to manage multiple GPUs inside a cluster. Whenever a GPU ends, its computation is reallocated with a new unaccomplished energy bin.

\subsection{Code modularity}

The code is natively developed with a modular approach, meaning that each section of the underlying model and related functions are organized in libraries which can be changed or integrated, with which ever exotic model one desire to test. The only requirement is that it fits the computational algorithm scheme illustrated in section \ref{sec:algorithm}. In fact, the code optimizations, illustrated in the following, are generalizable to other physical models and SDE expressions. As two starting examples, we provide in the code repository, a trivial empty model and a 1D COSMICA model. The first can be taken as the minimal algorithm structure shell with a trivial radial unphysical propagation. The letter, instead, contains the COSMICA general model, considering only radial particle propagation \cite{DellaTorre2016_OneD} and could be used as a preliminary test of agreement with the data.

\section{Code profile}

In this section, we profile the code during a standard simulation execution to disclose its efficient operations and bottlenecks. To perform this, we extract partial execution time of code sections and leverage the Nsight tools provided by Nvidia SDK.

We profile the latency of the various sections of the code and the single operations both for CPU and GPU APIs. An example of how the execution time progresses along a run is illustrated in the Table \ref{table:ex_time}, which refers to the same simulation configuration of Fig.~\ref{fig:A30_A40}. One can note that the total runtime is nearly completely compounded by the computations involving the stochastic propagation of the \textit{quasi-particles}. The latter consideration can be generalized to runs with all the energy bins, modulus the fact that little discrepancies of the proportion between section execution time can occur. This is due to the fact that the cycles on the energy bins, initial positions and stochastic propagation steps do have not a one-to-one correspondence. However, the predominance of the propagation section is expected to increase with multiple energy bins simulated.
Different outputs may result from other particle elements simulated. However, the disproportion of execution time between code sections is so deep that the main behaviour is not expected to change, not even with large-scale simulation in a larger cluster.

\begin{table}[htb!]
    \label{table:ex_time}
    \caption{Example of each code segment execution time}
    \begin{center}
    \small
    \begin{tabular}{| l | c |}
    \hline
        \multicolumn{1}{|c|}{\textbf{Code section}} & \textbf{Execution time (ms)} \\
        \hline
        \text{Back-propagation: Initialization} & 0.06 \\
        \hline
        \text{Back-propagation: propagation phase} & 3911.77 \\
        \hline
        \text{Back-propagation: Find Max} & 0.33 \\
        \hline
        \text{Back-propagation: Binning} & 0.15 \\
        \hline
        \text{Time to Set Memory} & 12.0 \\
        \hline
        \text{Time to create Rnd} & 0.1 \\
        \hline
        \textbf{Total} & 3924.4 \\
        \hline
    \multicolumn{2}{l}{The log file output shown here refers to a single simulation run} \\
    \multicolumn{2}{l}{for 1 energy bin, He3 isotope, from 19/05/2011 to 15/11/2017} \\
    \multicolumn{2}{l}{at the Earth position.}
    \end{tabular}
    \end{center}
\end{table}

\begin{figure}[htb!]
    \centering
    \includegraphics[width=1.0\linewidth]{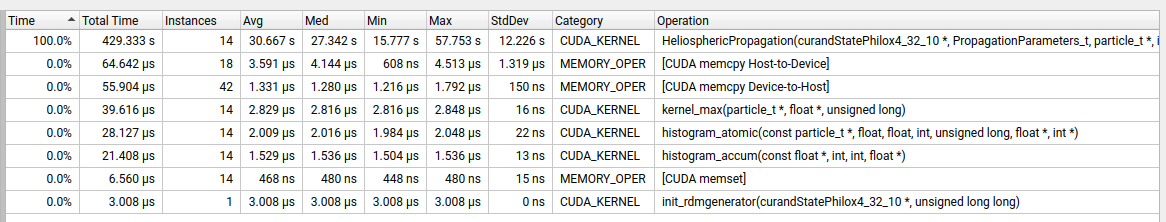}
    \caption{Example of the Nsight Compute profiling, executed on a single GPU. Here a single simulation is taken into account, for the same sample of Table \ref{table:ex_time}, but for all the energy bins.}
    \label{fig:A40_API}
\end{figure}

A deeper analysis of the GPU APIs execution is visible in Fig.~\ref{fig:A40_API}, where there is the Nsight System tool profiling of COSMICA run.
It confirms the preliminary conclusions derived from the analysis of Table \ref{table:ex_time}:
a) the memory copy and set operations are negligible as it is for the GPU's global memory allocation,
b) the computation related to energy histograms and random number generation requires less than $0.1\%$ of total run time
c) the function requiring nearly the totality of run time is the one unclosing heliospheric \textit{quasi-particle} propagation and local coefficients of SDE.
This is a remarkable result, because COSMICA is not affected by the most common bottlenecks of GPU parallel programming, and only a deeper optimization of the raw computations is needed in future versions. This is crucial, even, for scalability, because COSMICA can easily be distributed on a larger cluster without loss of efficiency: each GPU can manage an energy bin independently, and no expensive interactions with the CPU are required. This approach has no scaling limit for a given GPU cluster: if they are connected to a manager machine handling their IDs, the simulation is fractionated and distributed along all the available GPUs.

\section{Code optimizations}

In this section, we explore in detail the optimization procedures implemented in the code. We test each implementation separately and numbered from 1 to 6 in order to access their impact on performances.

\subsection{Version 1}

In this version was implemented the use of struct of arrays instead of array of structs, allowing synchronous broadcasting of memory access. Furthermore, Inside the propagation kernel, the \textit{quasi-particle} evolving variables are copied into a dynamic shared memory array till the propagation ends and the initial struct is updated.

\subsection{Version 2}

\begin{figure}[htb!]
    \centering
    \includegraphics[width=1.0\linewidth]{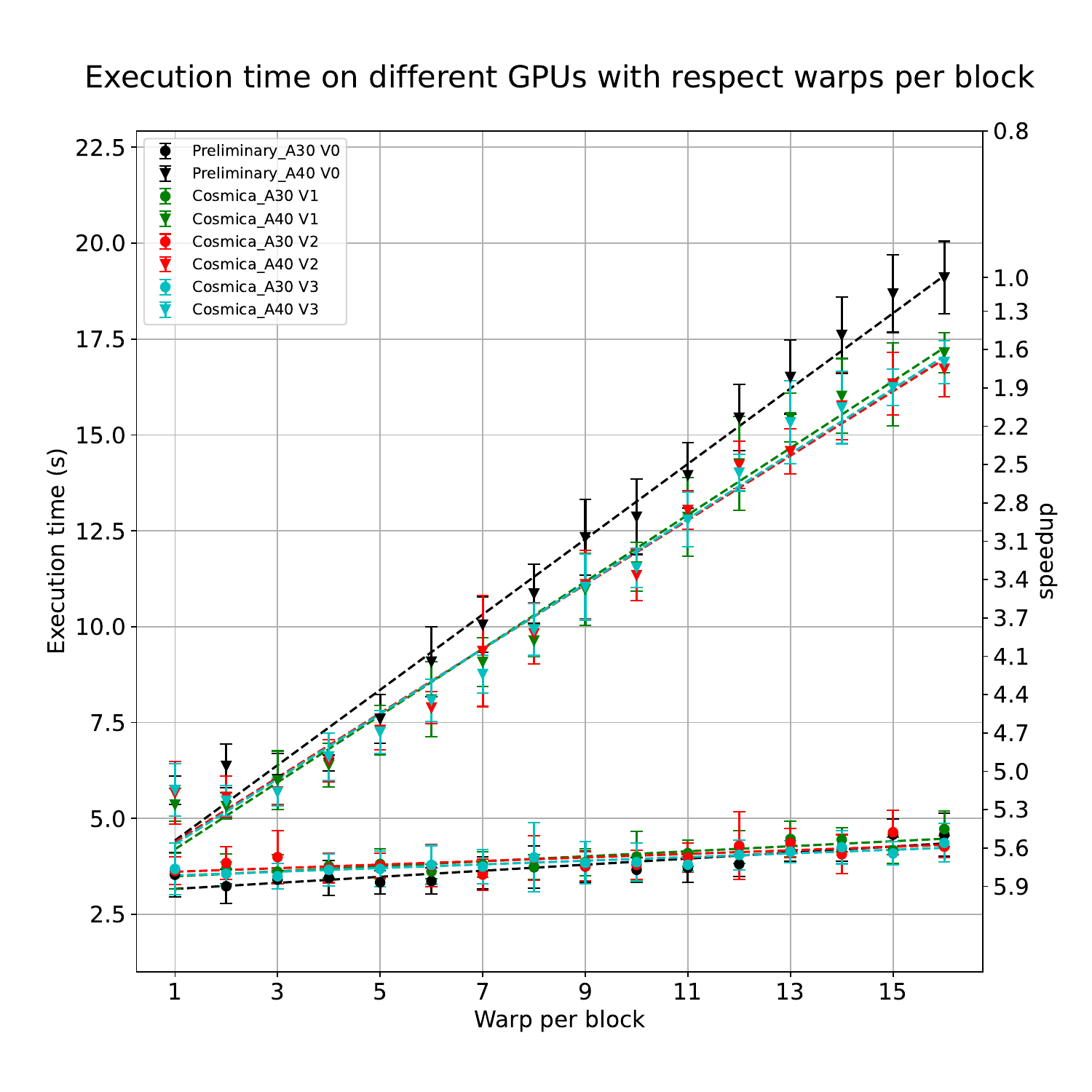}
    \caption{Execution time of the code versions 1-3 with respect to WpB allocated for each propagation kernel call. WpB are varied inside the GPU resources allocation boundaries (in particular registers). The error bars correspond to the standard deviation of execution time distributions of 10 identical COSMICA runs for each value of WpB. Circles (triangles) represent executions on the A30 (A40) board, for each version. The colors, instead, stance for the code version. The scale on the right axis shows the speed-up factor with respect to the maximum averaged execution time. This benchmark was performed simulating only 1 energy bin, with the He3 isotope, from 19/05/2011 to 15/11/2017 at the Earth position.}
    \label{fig:A30_A40}
\end{figure}

Here, we reduce the compiled part of the code for each chosen physical model for the simulation. This goes in the direction of modularity of the code. Meaning that different computational functions and physical models could be inserted in the same code structure, without burden the compilation and register pressure. Furthermore, we compute the optimal number of warps per block (WpB) to be allocated for the propagation kernel execution. This was done for the A30 and A40 Nvidia boards tested in our local cluster. The optimal WpB was extracted, running COSMICA on a test simulation and collecting the average execution time, varying the WpB from 1 to 16 which are the boundaries of cuda capabilities of the used boards. The results of this study can be seen in Fig.~\ref{fig:A30_A40}. Here we report the results for only the first three versions of the code, on which the analysis was centered. Nearly identical results are encountered in the subsequent versions.

As one can see, the execution time is strongly effected by the WpB on the A40 board, with a linear dependency. On the A30, instead, it has a lower slope, even if it presents an overall 1.5 speed up. From this analysis, the best WpB results to be equal to 2, for both the boards. Another crucial point one can note here, is that the A30 board is between 2 to 5 times faster than the A40 board, depending on the WpB value. This is unexpected with a preliminary analysis, because the A40 has a speeder clock, the same architecture and comparable resources, except for subdivision of warps and multiprocessors, with respect to the A30. The A40 boards have 84 multiprocessors, with 48 maximum warps each, with respect to 56 multiprocessors and 64 maximum warps of the A30 board. This could explain both the difference in execution time, between the two boards, and its strong WpB dependence in the A40. In fact, considering that our code saturates the registers per thread resource, this happens earlier in the latter board. Another explanation, could reside in the smaller memory bandwidth of the A40, which slow all the memory transfer contained inside the propagation kernel computations. A more detailed comparison of the board hardware can be seen in Table \ref{table:A30_A40}.

\begin{table}[htb!]
    \begin{center}
    \small
    \begin{tabular}{| l | c | c |}
    \hline
        & \textbf{A30} & \textbf{A40} \\
        \hline
        \textbf{Architecture} & Ampere & Ampere \\
        \hline
        \textbf{Boost clock speed} & 1440 MHz & 1740 MHz \\
        \hline
        \textbf{Core clock speed} & 930 MHz & 1305 MHz \\
        \hline
        \textbf{Peak FP32 performance} & 10.32 TFLOPS & 37.42 TFLOPS \\
        \hline
        \textbf{Maximum RAM} & 24 GB & 48 GB \\
        \hline
        \textbf{Memory bandwidth} & 933.1 GB/s & 695.8 GB/s \\
        \hline
        \textbf{Multiprocessors} & 56 & 84 \\
        \hline
        \textbf{Warps per Multiprocessor} & 64 & 48 \\
        \hline
    \end{tabular}
    \caption{NVIDIA's GPU boards comparison on the main hardware features of interest for the computations illustrated in this paper. All the values reported correspond to the maximum of the hardware available resources.}
    \label{table:A30_A40}
    \end{center}
\end{table}

\subsection{Version 3}

With this version of the code, we optimized the partial computations of the SDE coefficients and their mathematical expressions. In addition, we unpacked some variables meta-structures to reduce the register pressure along propagation kernel computations. This brings the registers per thread allocated from 96 to 80, which is a major improvement.

\subsection{Version 6}

This was a major update of the code, both for computational and scientific point of view. In this version, we passed from the energy based formulation of PTE (shown in \ref{eq:ParkerEquation}) to the momentum based one, which is the following:
\begin{align}
\label{eq:ParkerEquation_mom}
    \frac{\partial f}{\partial t}= &\frac{\partial}{\partial x_i} \left( K^S_{ij}\frac{\partial \mathrm{f} }{\partial x_j}\right) +\frac{P}{3}\frac{\partial V_{ \mathrm{sw},i} }{\partial x_i} \frac{\partial f}{\partial P} - \frac{\partial f}{\partial x_i} V_{ \mathrm{sw},i},
\end{align}
where $f \left(\overrightarrow{x}, p \right)$ is the omnidirectional distribution function, with p as particle momentum, and $P$ is the particle rigidity. The latter is defined as $P = \frac{p c}{Z e}$ . $f$ is related to the flux with the equation: $J = f P^2$ . In this frame, one passes from
\begin{align}
\label{eq:dR_L_mom}
    A_T &= \frac{2}{3}\frac{\alpha T V_{ \mathrm{sw}}}{r}, \quad to \quad A_P = \frac{2}{3}\frac{V_{ \mathrm{sw}}}{r} P,
\end{align}
and, most importantly, $L = 0$. The latter means that there is one less SDE to be solved. This reduces the computations needed at each propagation step and speed up the code. Furthermore, the CR propagation in heliosphere depends on rigidity and not kinetic energy as natural physical quantity. This translates in simpler calculations and in avoidance of redundant mathematical passages. The other components of $A$ and $B$, instead, remain the same, except for their reparametrization in function of particle rigidity instead of kinetic energy.


\section{Performances}

The performance comparison between the preliminary version of the CPU+GPU COSMICA code and the previous CPU-only implementation are reported in \cite{BOSCHINI20244302}. There one can see, also, how they scale with respect to the number of \textit{quasi-particle} simulated ($\mathcal{N}$). With it equal to $5 \cdot 10^4$, which is the typical value for a standard sample of simulation, the speed-up is of the order $40X$.

\begin{figure}[htb!]
    \centering
    \includegraphics[width=1.0\linewidth]{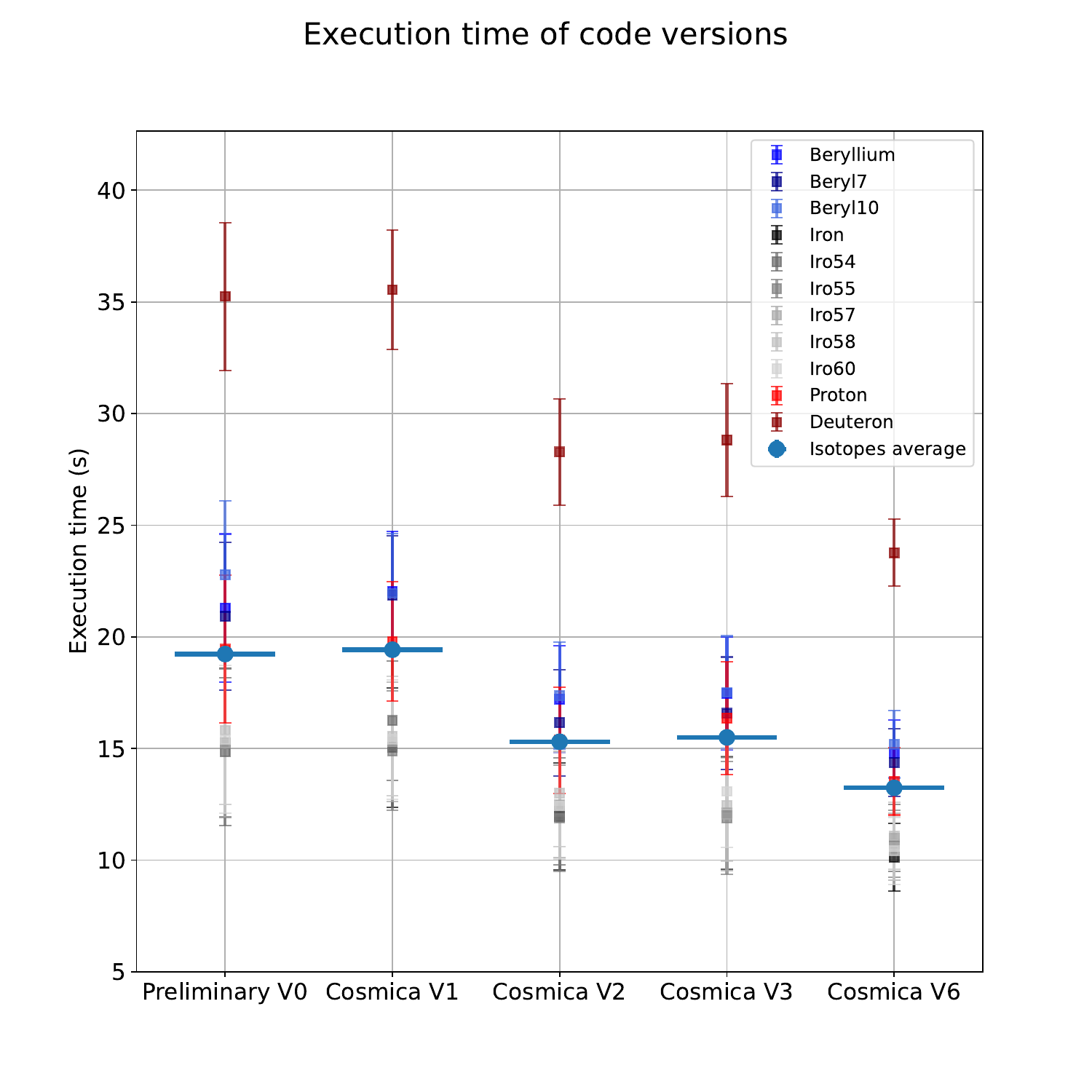}
    \caption{Comparison of the execution time of available stable code versions. The code was tested on hydrogen, beryllium and iron isotopes, present in COSMICA archive respectively indicated with blue, grey and red squared. The error bars in the plot correspond to standard deviation of execution time distributions of 10 COSMICA runs for each simulation configuration. The blue horizontal bars are placed at the average execution time along the ion partial averages.}
    \label{fig:version_speed}
\end{figure}

The performances of the illustrated versions are reported in Fig.~\ref{fig:version_speed}. The tests were performed simulating 11 checkpoint energy bins, with the proton and deuterium isotopes, from 19/05/2011 to 26/11/2013; with the Be7, Be9, Be10 isotopes, from 19/05/2011 to 26/05/2016; with the Fe54, Fe55, Fe56, Fe57, Fe58, Fe60 isotopes, from 19/05/2011 to 01/11/2019; all at the Earth position. There is clear the optimization trend of the code versioning. In fact, each version brings a speed-up or a constant behaviour within statistical error, with respect to the previous one. The two major performance jumps occur between version 1 and 2, and between version 3 and 6. This behaviour is expected, because Wpb optimization and rigidity formalism change are core modification, the first of the code execution scheduling, the second one of the propagation computations.
Moreover, the execution time of ions of the same elements are clustered, while they are considerably different for different elements. This is due to two factors: the propagation is strongly affected by the ion species at equal input energy, and the simulation periods taken into account have different dimension. Nevertheless, this is not an issue, but a feature inserted in the test set configurations to cover a representative set of possible simulations typically run by COSMICA. Overall, the last COSMICA version achieved a speed-up of about a factor $2X$, which results in a factor about $80X$ with respect to the CPU only implementation.

\section{Code validation}

\begin{figure}[htb!]
    \centering
    \includegraphics[width=1.0\linewidth]{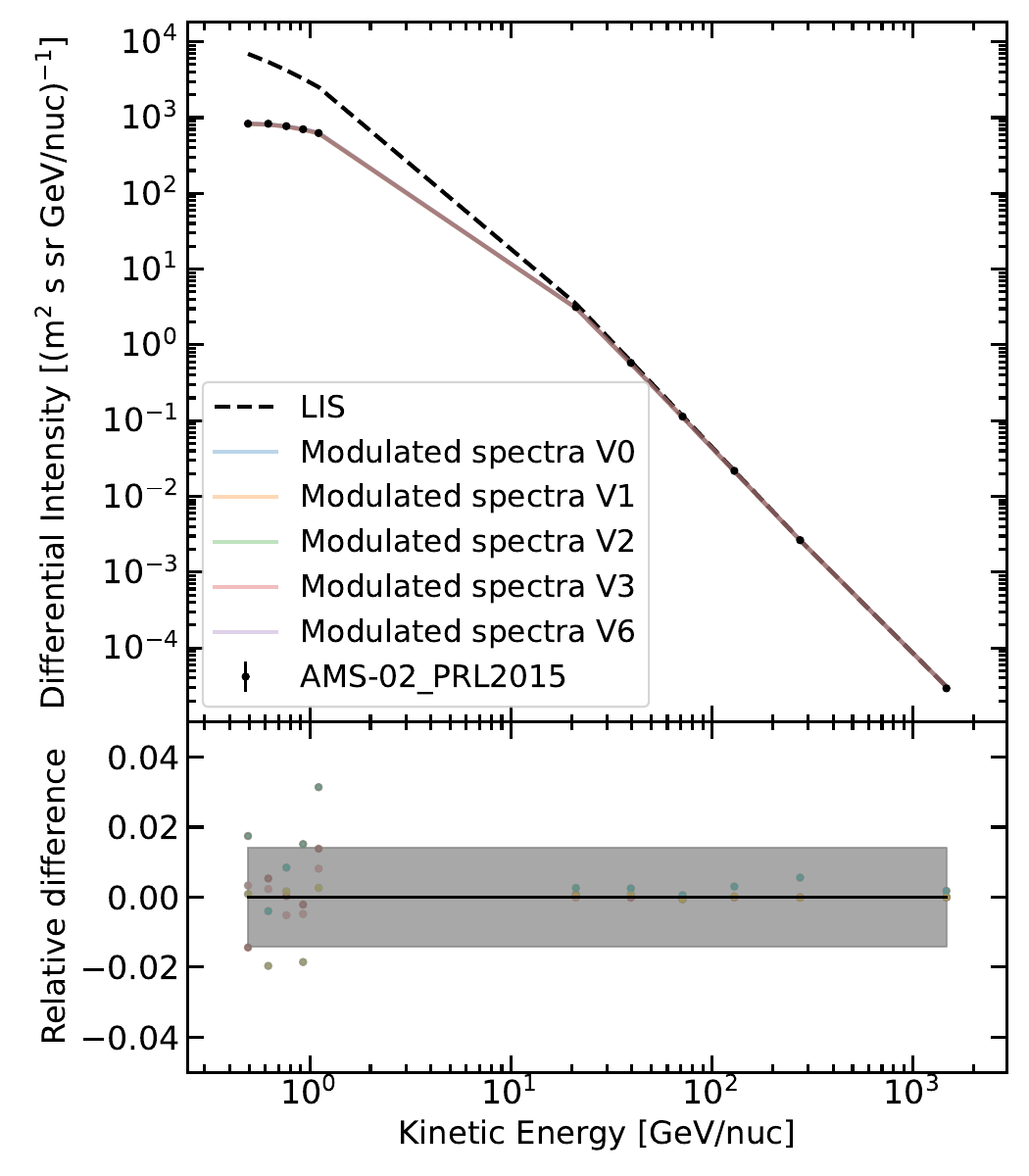}
    \caption{Upper panel: modulation of the protons spectra for the various code versions. The dashed line indicates the Galprop LIS. The black dots are, instead, the AMS-02 experimental data. The solid lines are the modulated spectra for the corresponding code version.
    Lower panel: discrepancies of the simulated modulated spectra for each code version with respect to the version 0. The scatter points represent the difference between version n and version 0. The grey band instead represent the statistical error of simulations calculated as $\frac{1}{\sqrt{\mathcal{N}}}$. Here we simulated 11 checkpoint energy bins, with the proton and deuterium isotopes, from 19/05/2011 to 26/11/2013 at the Earth position.}
    \label{fig:code_validation}
\end{figure}

To complete the picture of COSMICA is necessary to validate the code results with respect to experimental data and the version stability. Both purposes are fulfilled by the analysis illustrated with Fig.~\ref{fig:code_validation}. On its upper panel, one can check the goodness of the COSMICA simulations, which overlap with the AMS-02 data within the experimental error. In the lower panel, instead, we compare the modulated spectra of each code version, with respect to the preliminary version 0. As one can notice, the simulations produced by all the versions are within the stochastic statistical error associated to $\mathcal{N} = 5024$, overall. The larger discrepancies appear for small rigidities, at which the modulation and the stochastic nature of CR's propagation path are larger and decrease lying inside the statistical code uncertainty band, which is consistent with the results of \cite{BOSCHINI20244302}. Nevertheless, the code stability along versions is acceptably for the purposes of astrophysical studies usually performed leveraging the COSMICA code.

\section{Conclusions}

This paper introduces COSMICA, a high-performance parallel GPU code designed for the simulation of Cosmic Ray (CR) propagation within the heliosphere, its base algorithm, optimization, empirical and stability validations. Solving SDEs equivalent to the PTE, COSMICA significantly accelerates computational performance compared to CPU-based implementations. This improvement, achieved through GPU optimization techniques and architecture based optimization, enables a $\sim 80X$ speed-up with respect to CPU implementation, while maintaining the precision of results.
Key contributions of this work include the optimization of memory access patterns, the  wpB board  custom optimization, the modularization of the computational model, and the transition to a momentum-based formulation of the SDEs. These innovations reduce computational overhead and improve scalability, as demonstrated by the code's ability to efficiently distribute simulations across multi-GPU clusters.
Performance benchmarks validate the robustness and efficiency of COSMICA, showcasing speed-ups of up to 1.5X between successive optimized versions. The accuracy of the code is corroborated through its alignment with experimental data from AMS-02, establishing its suitability for astrophysical research and space mission planning. An example of researches, available with these simulation capabilities, is the systematic and wide-range model parameter research, which is executed by average the multiple simulation fittings on experimental data.

\section{Acknowledgment}

This activity is supported by Fondazione ICSC , Spoke 3 Astrophysics and Cosmos Observations. National Recovery and Resilience Plan (Piano Nazionale di Ripresa e Resilienza, PNRR) Project ID $CN_00000013$. MG, SDT and GLV are supported by INFN and ASI under ASI-INFN Agreement No. 2019-19-HH.0 and its amendments.



\end{document}